\documentclass[preprint2]{aastex}

\def\msun{$M_{\odot}$}

\def\ergcm{\hbox{erg cm$^{-2}$ s$^{-1}$ }}

\def\xte{{\it RXTE~}}
\def\h1743{H1743--322}

\slugcomment{}

\begin{document}

\title{The X-ray Outburst of \h1743: High-Frequency QPOs with a 3:2 
Frequency Ratio}


\author{Ronald A. Remillard}
\affil{Center for Space Research, MIT, Cambridge, MA 02139-4307 }
\email{rr@space.mit.edu}

\and

\author{Jeffrey E. McClintock}
\affil{Harvard-Smithsonian Center for Astrophysics, 60
Garden St. MS-3, \\ Cambridge, MA 02138}
\email{jem@cfa.harvard.edu}

\and

\author{Jerome A. Orosz}
\affil{Dept. of Astronomy, San Diego State University, 5500 Campanile 
Drive, \\ San Diego, CA 82182-1221}
\email{orosz@sciences.sdsu.edu}

\and
\author{Alan M. Levine}
\affil{Center for Space Research, MIT, Cambridge, MA 02139-4307 }
\email{aml@space.mit.edu}

\begin{abstract}

The 2003 X-ray outburst of the candidate black-hole binary, \h1743,
was investigated in frequent pointed observations (2--250 keV) with
the {\it Rossi} X-ray Timing Explorer.  We consider one particular
program of 130 observations.  We organized these data into 111 time
intervals and conducted a search for the presence of high-frequency
quasiperiodic oscillations (HFQPOs) in the range 50--2000 Hz.  Only a
single observation (2003 June 13) yielded a detection above $4
\sigma$; the central frequency of $239 \pm 4$ Hz is consistent with
the 240 Hz QPO reported for this source on 2003 May 28 \citep{hom03}.
We next grouped the observations in several different ways and
computed the average power-density spectra (PDS) in a search for
further evidence of HFQPOs.  This effort yielded two significant
results for those observations defined by the presence of
low-frequency QPOs (0.1-20 Hz) and an absence of ``band-limited''
power continua: (1) The 9 time intervals with the highest X-ray flux
yielded an average PDS with a QPO at $166 \pm 5$ Hz. ($4.1 \sigma$;
3--35 keV); and (2) a second group with lower X-ray flux (24 time
intervals) produced an average PDS with a QPO at $242 \pm 3$ Hz ($6.0
\sigma$; 7--35 keV).  The ratio of these two frequencies is $1.46 \pm
0.05$.  This finding is consistent with results obtained for three
other black hole systems that exhibit commensurate HFQPOs in a 3:2
ratio.  Furthermore, the occurrence of H1743-322's slower HFQPO at
times of higher X-ray luminosity closely resembles the behavior of
XTE~J1550-564 and GRO~J1655-40.  We discuss our results in terms of a
resonance model that invokes frequencies set by general relativity for
orbital motions near a black-hole event horizon.

\end{abstract}


\section{Introduction}

A key question in black hole binary (BHB) research is whether
high-frequency QPOs (HFQPOs), observed in the range of 40--450 Hz,
represent a unique timing signature that may constrain a black hole's
mass and spin via a model rooted in general relativity (GR; see
McClintock \& Remillard 2003). \nocite{mcrem03} The HFQPO signals are
weak, and an attack on this problem requires observations of bright
X-ray transients during their rare and short-lived ($\sim$ months)
outbursts.

HFQPOs have been detected in 7 confirmed or candidate BHB systems,
including \h1743 (see below). In each of three sources (GRO~J1655--40,
XTE~J1550--564, and GRS~1915+105), transient HFQPOs have appeared at a
pair of frequencies that are commensurate with a 3:2 ratio
\citep{rem02a,rem03}.  Furthermore, when the HFQPOs were compared
among the three sources, it was found that the frequencies scale
inversely with black hole mass. This result is consistent with
expectations for oscillations produced at some characteristic radius
in a strong-gravity environment described by GR theory, but
only in the special case where the values of the dimensionless black
hole spin parameter are similar for the three sources \citep{mcrem03}.

Commensurate HFQPO frequencies can be interpreted as a signature of an
oscillation driven by some type of resonance condition.  Abramowicz \&
Kluzniak (2001) \nocite{abr01} had earlier proposed that QPOs could
represent enhanced emission from a particular radius where there is a
resonance in the GR coordinate frequencies for orbital/epicyclic
motions in strong gravity \nocite{mer99} (see Merloni et al. 1999).
Resonances in some form may be applicable to both BH and NS systems
\citep{abr03}. We note that GR coordinate frequencies and associated
beat frequencies in the inner accretion disk were invoked in earlier
work on variable-frequency QPOs in both neutron-star and some BHB
systems (Stella, Vietri, \& Morsink 1999). \nocite{ste99}

Herein, we report on a general search for HFQPOs in the source \h1743.
The 2003 outburst of this source was first detected on March 21 in
hard X-rays (15--200 keV) by INTEGRAL (IGR~J17464--3213; Revnivtsev et
al. 2003). \nocite{rev03} Follow-up observations with \xte led to the
recognition that the source is a recurrent X-ray nova first observed
with HEAO1 in 1977--1978 \citep{mar03,gur78}. The X-ray spectral and
temporal properties exhibited during the first outburst had
established \h1743 as a black-hole candidate; its X-ray spectrum
contained both a soft component (1--10 keV) and a hard X-ray tail
(10--100 keV; Cooke et al. 1984), \nocite{coo84} and there were no
pulsations or X-ray bursts that would identify the source as a
neutron-star system \citep{tan95}.

An X-ray light curve (1.5--12 keV) for the recent outburst is
displayed in the top panel of Fig. \ref{fig:asm}, using data from the
\xte All-Sky Monitor (ASM; Levine et al. 1996). \nocite{lev96} The
bottom panel shows measurements of the spectral hardness ratio,
defined as the ratio of the ASM count rate at 5--12 keV to that at
3--5 keV. These data show that the initially hard spectrum rapidly
evolves to a softer one. A number of bright flares are seen
before the source settles into a relatively steady and soft X-ray
state.  Several bright flares reach intensities near 1.5 Crab
(i.e. $3.6 \times 10^{-8}$ \ergcm at 2--10 keV) during 2003 April
18-24.

Radio detections of \h1743 were obtained during the first few weeks of
the 2003 outburst, yielding a precise celestial position and evidence
for a transient radio jet (Rupen, Mioduszewski, \& Dhawan 2003a;
2003b; 2003c).  \nocite{rup03a,rup03b,rup03c} Soon thereafter, a faint
optical counterpart was detected (R=21.9), despite $\sim 13$ mag of
dust extinction \citep{ste03}.

\xte made 227 pointed observations of \h1743 (=XTE~J1746--326) during
the 2003--2004 outburst, in the course of carrying out five different
observing programs.  In this paper we report on the results of a
search for HFQPOs (50--2000 Hz) from this source using the 130
observations from program ID 80146. The times of these observations
are indicated in the row of vertical lines in the top panel of
Fig. \ref{fig:asm}.  An HFQPO was reported for \h1743 after a 25 ks
observation with \xte on 2003 May 28 for program ID 80135
\citep{hom03}.  This observations yielded an HFQPO detection ($4.5
\sigma$) at 240 Hz, with a possible second feature ($2.5 \sigma$) at
160 Hz \citep{hom03}.

In the following sections we describe our analysis techniques and
methods for data grouping that lead to significant detections of
HFQPOs at 242 and 166 Hz.  Assuming \h1743 is a genuine BH, our
discovery establishes this source as the fourth BH where a pair of
HFQPOs appear to have commensurate frequencies in a 3:2 ratio.  It
is further be shown that additional properties of these HFQPOs
bear a striking resemblance to those seen in GRO~J1655--40 and
XTE~J1550--564 \citep{rem02a}.

\section{Observations and Data Analysis}

We consider all of the 130 observations of \h1743 conducted by \xte
under program ID 80146.  This data set contains occasional sequences
of short observations, and we therefore chose to collect and organize
the observations into the 111 time intervals that are listed in Table
\ref{tab:obs}.  The number of observations combined for each time
interval is given in column 4. The maximum time span for any interval
is 0.48 days, and the total exposure time is 498 ks.

Our X-ray timing analyses utilize data from the \xte~Proportional
Counter Array (PCA).  The data telemetry modes chosen for this program
allow us to construct PCA light curves with $125 \mu$s time resolution
in 3 energy bands: 3-7, 7-14, and 14-35 keV.  The PCA consists of 5
detector units (PCUs).  However, the observations of \h1743 utilized a
variable number of PCUs (usually 3) because some of the PCUs are
cycled off periodically to avoid problems with detector breakdown.
For each observation interval, the telemetry modes for high-speed data
merged all of the good events from the PCUs that were in operation.

Table \ref{tab:obs} also provides, for each observation interval, the
mean count rate (c/s/PCU) at 2-35 keV (col. 6) and at 7--35 keV
(col. 7). The latter quantity is useful in evaluating the strength of
nonthermal radiation in \h1743 and was used to group observations in
the search for HFQPOs. The PCA count rates are derived from
background-subtracted energy spectra processed from the `Standard 2'
data mode using PCU \#2, which was in use (along with PCU \#0) during
every observation of program 80146. All of the spectral reductions
were performed via the `ftools' routines in the HEAsoft software
package distributed by NASA's HEASARC.

The procedures used to compute power density spectra (PDS) and to
search for QPOs are described in Remillard et al. (2002).
\nocite{rem02a} Briefly, for each of the 111 time intervals (Table
\ref{tab:obs}), we compute Fourier transforms for each 256-s segment
in the PCA light curve, and we output the average PDS for the
available number of segments. We then subtract the deadtime-corrected
Poisson noise and normalize the PDS to units of (rms deviation /
mean)$^2$ Hz$^{-1}$.  In each frequency bin, we compute the
uncertainty as the larger of either the statistical error ($2/N^{0.5}$
prior to rms normalization) or the empirical standard deviation of the
mean power, where $N$ is the number of transforms (of 256 s duration)
in a given observation interval. We note that this conservative
approach can yield error bars that appear large, compared to
bin-by-bin fluctuations, when the broad power continuum varies
significantly during an observation. Finally, the power densities and
uncertainties are re-binned in logarithmic intervals of frequency
($\nu$), maintaining (for this study) a minimum $\Delta \nu / \nu =
0.04$.  When we average the PDS for groups of observations, we
compute the weighted mean (using $\sigma^{-2}$) and the net uncertainty
for each frequency bin.

We search for HFQPOs using statistical tests to detect a Lorentzian
peak rising above the local power continuum. The continuum, in terms
of log $P_{\nu}$, is modeled with a second order polynomial in log
$\nu$, which presumes a power-law function with allowances for broad
curvature.  QPOs are distinguished from broad power peaks using a
coherence parameter, $Q = \nu / FWHM \ga 2$. For each PDS, we use
$\chi^2$ minimization to obtain the best fit for the QPO profile and
the local power continuum. Finally, in our general QPO searches, we
seek results that have acceptable values of $\chi_{\nu}^2$ while
surpassing a significance threshold for the integrated QPO power
($P$), relative to the uncertainty ($\sigma_P$): $P/\sigma_P \ga 4$.
This elevated threshold compensates for the large number of PDSs and
trial frequencies considered in the analysis of data sets obtained for
typical bright X-ray transients.

\section{Results}

We searched for QPOs in the range 50--2000 Hz for each of the 111
observing intervals (Table \ref{tab:obs}).  For each interval, we
considered separately the average PDS for the energy ranges 3-35,
7-35, and 14-35 keV. Only one interval (2003 June 13 = MJD 52803)
yields a detection above $4 \sigma$.  In this case the energy range is
7--35 keV, the central frequency is $239 \pm 4$ Hz, the integrated rms
amplitude is $r = 2.0 \pm 0.2$~\%, $Q \sim 9$, and the detection
significance is 4.3 $\sigma$.  The central frequency is consistent
with the 240 Hz QPO detected for this source on 2003 May 28 (MJD
52787) in observations from a different \xte program \citep{hom03}.
There is one additional noteworthy finding from this initial search
for HFQPOs.  For the interval with highest PCA count rate (1.5 Crab on
MJD 52765), an HFQPO candidate ($3.8 \sigma$) appears at $162 \pm 7$
Hz with $Q \sim 4$ and $r = 1.1 \pm 0.2$~\% (3-35 keV).

Most of the HFQPOs detections in BHB outbursts have been found in PDS
averaged over a number of \xte observations \citep{rem99,cui00,rem02c}
because high statistical precision is required in order to detect a
faint signal spread over a bandwidth of 20 to 50 Hz.  Furthermore,
judicious data selection is required, since HFQPOs are usually
detected only in the ``steep power-law state'' \citep{mcrem03}, and
within this state further considerations may be required to deal with
amplitude variability, frequency switching, and choice of PDS energy
range. We therefore investigated several strategies for grouping the
observations of \h1743 prior to further searches for HFQPO detections.

In the case of XTE~J1550-564, it was shown that the phase lags (13-30
keV vs. 2-13 keV) in low-frequency QPOs (LFQPOs; 0.1-20 Hz) could be
used to define three LFQPO types that are well correlated with HFQPO
properties \citep{rem02c}. These LFQPO types predicted, respectively,
the absence of HFQPOs or the appearance of HFQPOs at either 185 or 276
Hz. We investigated this method for the case of \h1743, but we found
the results to be statistically unsatisfactory because the source is
fainter than XTE~J1550-564 and the LFQPO amplitudes are 
comparatively low.

As an alternative approach, one can capitalize on correlations found
between HFQPOs and the properties of the power continuum and the
energy spectrum.  The LFQPO type (``C'') that forecasts an absence of
HFQPOs in XTE~J1550-564 can also be recognized for its association
with a ``band-limited'' power continuum in which the power density is
flat at low frequencies and then drops abruptly at frequencies above the
LFQPO and its harmonic. Such a PDS can be recognized via the
integrated $rms$ power derived from the PDS (0.1 to 10 Hz), and those values
are given in col. 8 of Table \ref{tab:obs}. Observation intervals for
\h1743 with integrated $rms > 0.11$ all show LFQPOs and band-limited
power continua, and we label these as PDS type ``q-bl'' in col. 9 of
Table \ref{tab:obs}.  Intervals with PDS that are devoid of LFQPOs are
noted as PDS type ``0''; these cases also exhibit the lowest
values of the PCA hardness ratio ($HR = $col. 6 / col. 7 $ < 0.017$,
with one exception). These properties suggest that PDS type 0
designations identify the times when the source is in a
thermal-dominant state \citep{mcrem03}.  The remaining intervals
contain LFQPOs without band-limited power continua and are labeled
``q'' in col. 9 of Table \ref{tab:obs}.

We computed average PDS for groups ``q-bl'' (23 observation intervals)
and ``0'' (44 intervals), excluding the
last 9 intervals where the 2-35 keV count rate drops below 550
c/s/PCU ($\sim 0.2$ Crab) and the PDS become statistically weak. We
then divided the ``q'' type PDS into 2 groups distinguished by the
source brightness above or below 430 c/s/PCU at 7-35 keV (col. 8 of
Table \ref{tab:obs}).  This latter criterion attempts to exploit the
fact that the strength of the steep power-law component distinguishes
the observations that yield each of the QPOs with frequencies that scale
in a 3:2 ratio (hereafter referred to as the $2 \nu_0$ QPO and the $3 \nu_0$ 
QPO) in both XTE~J1550-564 and GRO~J1655-40 \citep{rem02a}.

\begin{figure}

\figurenum{2} \plotone{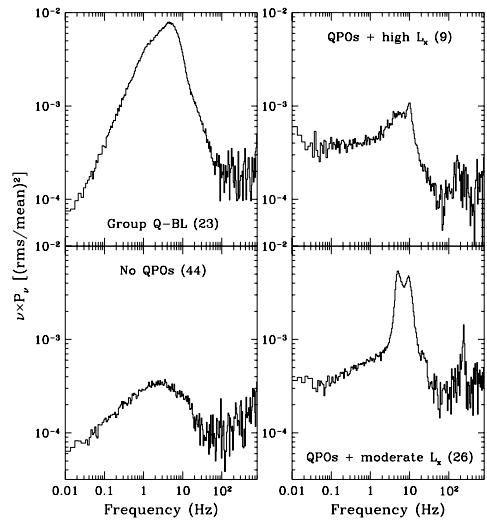}
\caption{Average PDS, in units of log ($\nu \times P_{\nu}$), for
observations of H1743-322 grouped by the properties of the power
continua and the X-ray flux. The power spectra are computed for
photons in the range 2-35 keV, except for the group in the
bottom-right panel, where the PDS at 7--35 keV is shown.  HFQPOs are
not seen when there is a band-limited power-continuum (group ``Q-BL'';
see text), nor when LFQPOs are absent (during times that correlate
with very soft spectra). However, HFQPOs are seen (right panels) when
LFQPOs are present and the band-limited power continuum is diminished.
Furthermore, the HFQPOs appears to switch between two commensurate
frequencies when these cases are divided into two ranges of luminosity
above 7 keV. \label{fig:pdstypes}}

\end{figure}

The PDS for the 4 groups, plotted in units of log($\nu \times
P_{\nu}$) vs. log $\nu$, are shown in Fig. \ref{fig:pdstypes}.  The
PDS fro the ``q-bl'' group shows a strong and very broad feature that
peaks near 3 Hz.  The PDS for the ``0'' group appears to show the same
feature, although reduced in strength by a factor $\sim 30$.  HFQPO
features are apparent only in the PDS for the ``q'' groups, and they
occur at different frequencies.  We note that all of the four PDS show
a residual continuum above 100 Hz that is an artifact of an imperfect
deadtime model for the PCA instrument.

\begin{figure}

\figurenum{3} \plotone{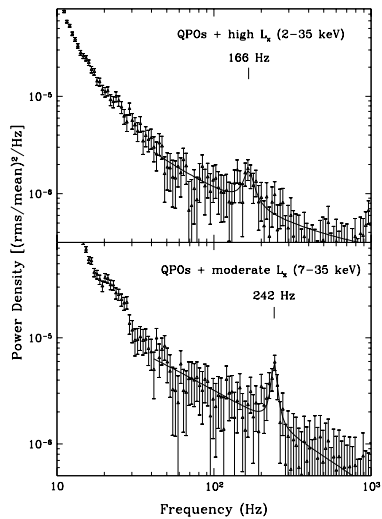}
\caption{Magnified view of the high-frequency QPOs for the same two 
groups that have their average PDS displayed in the right panels of Fig. 2. 
The solid lines show the QPO profile fits, which assume a Lorentzian 
profile above the local power continuum.  The fits are computed and 
displayed in units of $P_{\nu}$, rather than the units of $\nu \times P_{\nu}$
used in Fig. 2.  The central frequencies are $166 \pm 5$ Hz for the high 
$L_x$ group at 2-35 keV, and $242 \pm 3$ Hz for the lower $L_x$ group at 7-35 
keV. These results are statistically consistent with an interpretation
as commensurate frequencies scaled in a 3:2 ratio. \label{fig:pdsmod} }

\end{figure}

In Fig. \ref{fig:pdsmod} we show the HFQPO detections for the two
``q'' groups at higher frequency resolution, along with the fitted
models for the QPO profiles.  The 9 ``q'' intervals with the higher
X-ray fluxes produce an average PDS (2-35 keV) with a QPO ($4.1
\sigma$) at $166 \pm 5$ Hz, with $r = 0.60 \pm 0.08$ \% and $Q = 5.7
\pm 1.6$.  The ``q'' group comprising 26 intervals at lower X-ray flux
produces an average PDS (7-35 keV) with a QPO ($6.0 \sigma$) at $242
\pm 3$ Hz, with $r = 1.1 \pm 0.1$ \% and $Q = 11 \pm 1$.  Note that
the transition in \h1743 from a broader HFQPO at $2 \nu_0$ in a broad
bandwidth to a narrower feature at $3 \nu_0$ in a harder energy band
occurs with decreasing 7-35 keV count rate, which is presumably
dominated by a non-thermal spectral component.  This is precisely the
behavior exhibited by the HFQPOs from XTE~J1550-564 and GRO~J1655-40
\citep{rem02a}.

\section{Discussion}

The X-ray light curve and variability characteristics of \h1743 during
its 2003 outburst fully support the identification of this source as a
black hole candidate (see \S 1).  The behavior of \h1743 resembles the
BHBs XTE~J1550-564 and GRO~J1655-40 in many ways, although it was not
as bright as those sources (by a factor of 3--4), comparing the
brightest 4-month intervals of each outburst.  This is the likely
reason why we do not detect HFQPOs from \h1743 during individual
observations, other than those on MJD 52787 and possibly MJD 52765.

The effort to group the PDS in order to gain statistically significant
HFQPO detections for \h1743 was guided by results for XTE~J1550-564
and GRO~J1655-40 \citep{rem02a}. Our best results were found simply by
selecting observations with LFQPOs present, excluding cases with q-bl
type power continua, and then grouping the PDS according to the 7-35
keV count rate, which represents the strength of the nonthermal X-ray
flux.  This effort yielded two HFQPOs with properties that closely
resemble the HFQPOs of XTE~J1550-564 and GRO~J1655-40, i.e. the
central frequencies are consistent with a 3:2 ratio, the amplitudes
have $rms \sim $ 1\%, and the lower frequency QPO (166 Hz for \h1743)
is seen at times of highest non-thermal flux.  In addition, the higher
frequency QPO (242 Hz) has a narrower profile (i.e. higher $Q$ value)
and the detection bandwidth has a relatively higher mean photon
energy. We note that the one remaining BH system with HFQPOs in a 3:2
ratio (GRS~1915+105) was investigated using a different analysis
method in which each QPO was extracted from a portion of a particular
type of violently variable light curve \citep{rem02b,rem03}.

\begin{figure}

\figurenum{1} \plotone{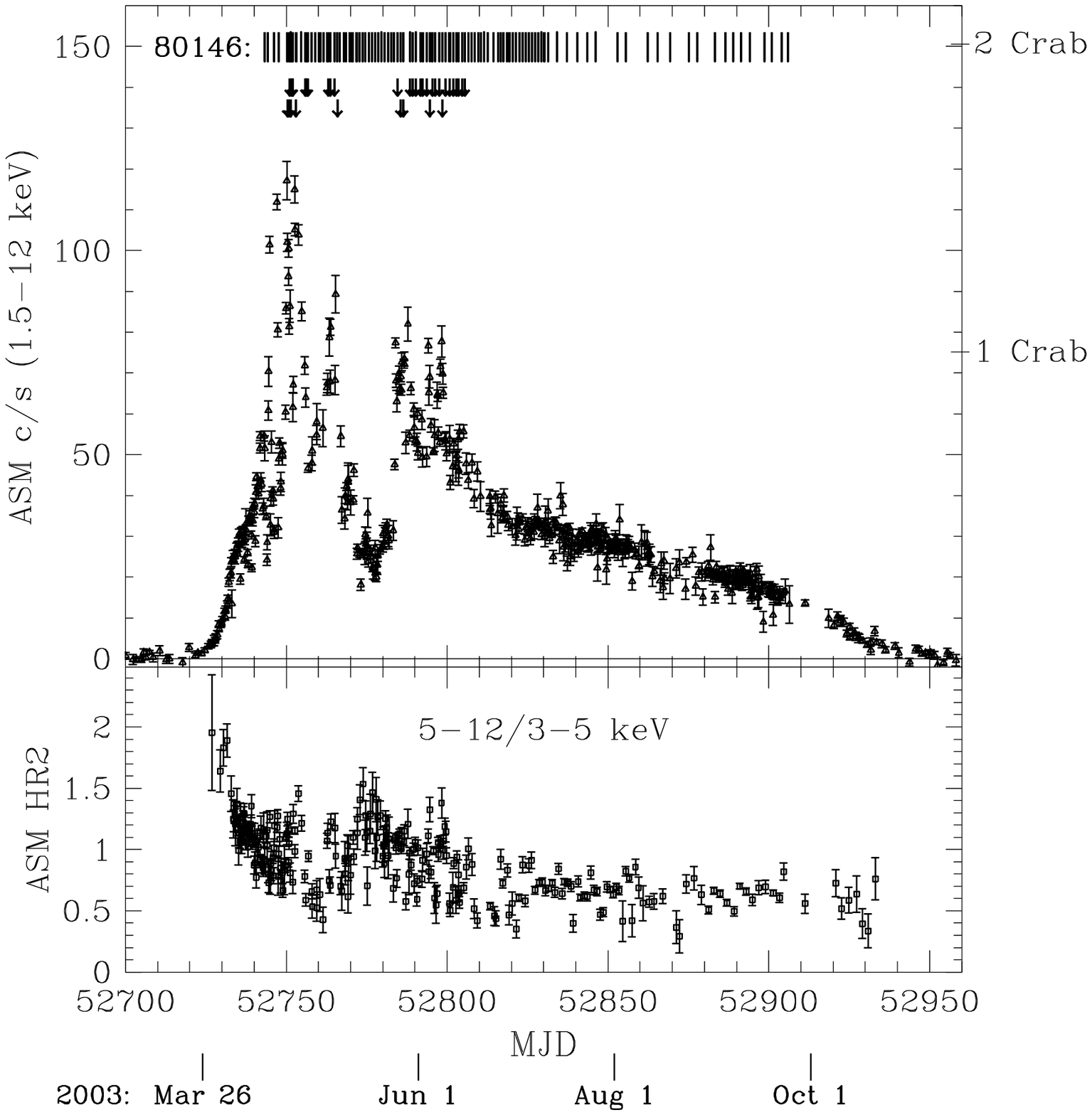}
\caption{\xte ASM light curve (1.5-12 keV) and hardness ratio (HR2)
covering the 2003 outburst of H1743-322. For reference, the count rate
for the Crab Nebula is 75.5 ASM c/s. The tick marks in the top panel
indicate the times of RXTE pointed observations considered in this
paper. The arrows show the times of observations contributing
to QPO detections at 166 Hz (upper row) and 242 Hz (lower row),
as explained in \S 3. \label{fig:asm}}

\end{figure}

HFQPOs with frequencies at $2 \nu_0$ and $3 \nu_0$ can be considered
as expressions of a single frequency system. As with the other BHBs
discussed above, the frequency system in \h1743 appears to be invariant
through changes in luminosity or interruptions due to the source's
evolution through a thermal-dominant or other states.  The observations
that correspond with the two groups that yield HFQPO detections 
(i.e. PDS type ``q'' in Table 1) span a range 204--1083 c/s/PCU at
7--35 keV and 1234--3775 c/s/PCU at 2--35 keV.  The temporal order of
these observations is shown with small arrows in the top panel of
Fig. \ref{fig:asm}, where the upper/lower row represents the 242/166
Hz group, respectively.  There are three gaps in the series of q-type
observations (i.e. combined arrows) that occur when
the PDS types are either 0 or q-bl (see Table 1). 

In summary, black hole HFQPOs appear at stable pairs of frequencies
with 3:2 ratio, and they exhibit common behavior patterns linked to
the properties of the energy spectra. These results support the view
that HFQPOs convey a distinct temporal signature for each accreting
black hole. The detection bandwidth and high frequencies suggest that
HFQPOs originate near the black hole event horizon. Furthermore, the
stability of HFQPO pairs with changing luminosity suggests that the
frequencies may depend only on the inherent properties of the black
hole, viz. its mass and spin.

We have shown that the HFQPOs in \h1743 appear to originate from the
same mechanism as the HFQPOs with 3:2 frequency ratio in the other
BHBs. An empirical relationship has been derived (only 3 sources)
between the HFQPO frequency systems and black hole mass
\citep{mcrem03}, and we may apply this to the case of \h1743.  We
assume that HFQPO frequencies depend only on the black hole mass
($M_x$) and spin, which is frequently evaluated in terms of the
dimensionless spin parameter: $a_* = cJ/GM_x^2$, where $J$ is the
angular momentum of the black hole. If the black hole in \h1743 has a
similar value of $a_*$ as the other BHBs, then the HFQPOs (with $\nu_0
\sim 81$ Hz) suggests: $M_x = 931 / \nu_0 \sim$ 11.5 \msun.

Detailed models of the HFQPO properties and behavior patterns must
confront the problem that we do not understand the origin of the steep
power-law spectrum.  This radiation component is always fairly strong
when the 3:2 oscillations appear. Furthermore, the strength of the
steep power-law appears to regulate the frequency switching from $2
\nu_0$ to $3 \nu_0$. 

As noted in \S 1, commensurate HFQPOs in BHBs
have been interpreted with a ``parametric resonance'' concept that
hypothesizes enhanced emissivity from accreting matter at a radius
where two of the three coordinate frequencies (i.e. azimuthal, radial,
and polar) have commensurate values that match (either directly or via
beats) the observed QPO frequencies.  For the cases with optical
determinations of the black hole mass, the value of $a_*$ can be
determined via the application of this resonance model if the correct
pair of coordinate frequencies can be identified.  Reasonable values
($0.25 < a_* < 0.95$) are derived from the observed HFQPOs for either
2:1 or 3:1 ratios in either orbital:radial or polar:radial coordinate
frequencies \citep{abr01,rem02a}. The driving mechanism that would
allow accretion blobs to grow and survive at the resonance radius has
not been specified, and it is known that there are severe damping
forces in the inner accretion disk \citep{mark98}. On the other hand,
MHD accretion simulations under GR do show transient condensations in
the inner disk \citep{haw01}.  Ray-tracing calculations under GR
\citep{schn04} show that the putative blobs could indeed produce the
HFQPO patterns, and that the appearance       of $3~\nu_0$ versus
$2~\nu_0$ for the stronger QPO is governed by the angular width
of the blob.  Clearly, more work is needed to
investigate this resonance model.

An alternative scenario is to extend the models for ``diskoseismic''
oscillations to include non-linear effects that might drive some type
of resonant oscillation.  Diskoseismology treats the inner disk as a
resonance cavity in the Kerr metric \citep{kat01,wag99}.  Normal modes
have been derived for linear perturbations, and the extension of this
theory would be both very interesting and difficult.  There are also
models that invoke oscillations from geometries other than a thin
disk, e.g. the accretion torus of Rezzolla et al. (2003).
\nocite{rez03} In any case, BHB HFQPOs deserve careful study as a
potential opportunity to derive primary information about black holes
while developing astrophysical applications for the strong-field regime
of GR theory.

\acknowledgements Partial support for R.R. was provided by the NASA
contract to MIT for RXTE instruments. J.M. acknowledges partial
support from NASA grant NAG5-10813.  We thank Jeroen Homan for
productive discussions about H1743-322.

\clearpage

\begin{deluxetable}{rllcrrccc} 
\tablenum{1}
\tabletypesize \scriptsize
\tablecaption{RXTE Observations of H1743-322: Program 80146}
\tablehead{
\colhead{\#} & \colhead{RXTE\tablenotemark{a}} & \colhead{MJD\tablenotemark{b}} &
\colhead{Num.} & \colhead{Exposure} & \colhead{PCA Rate\tablenotemark{c}} &
\colhead{PCA Rate\tablenotemark{d}} & \colhead{rms\tablenotemark{e}} & 
\colhead{PDS} \\
\colhead{} & \colhead{Day} & \colhead{Time} & \colhead{Obs.} & 
\colhead{(s)} &
\colhead{2-35 keV} & \colhead{7-35 keV} & \colhead{(0.1-10 Hz)} & 
\colhead{Type} \\
}

\startdata
1   &  3390  &  52743.24  &  1  &   2679  & 1940.2  &  461.5  &  0.112  &  q-bl \\
2   &  3391  &  52744.22  &  1  &   3059  & 1550.1  &  398.4  &  0.137  &  q-bl \\
3   &  3393  &  52746.20  &  1  &   2910  &  708.8  &  193.0  &  0.151  &  q-bl \\
4   &  3394  &  52747.63  &  1  &   3206  & 1024.2  &  247.6  &  0.117  &  q-bl \\
5   &  3397a &  52750.31  &  2  &   5783  & 2320.3  &  608.2  &  0.050  &  q \\
6   &  3397b &  52750.81  &  1  &  15520  & 2061.4  &  495.2  &  0.101  &  q \\
7   &  3398a &  52751.09  &  2  &   6899  & 1660.9  &  384.3  &  0.101  &  q \\
8   &  3398b &  52751.32  &  2  &   8406  & 3478.3  &  882.2  &  0.085  &  q \\
9   &  3398c &  52751.71  &  1  &   3282  & 1763.3  &  365.6  &  0.104  &  q \\
10  &  3398d &  52751.99  &  1  &   3653  & 1694.8  &  326.9  &  0.089  &  q \\
11  &  3399  &  52752.97  &  1  &  10764  & 2281.3  &  575.7  &  0.077  &  q \\
12  &  3400  &  52753.17  &  1  &   2978  & 2255.7  &  596.4  &  0.048  &  0 \\
13  &  3401  &  52754.58  &  2  &   3542  & 1743.5  &  383.4  &  0.117  &  q-bl \\
14  &  3402  &  52755.96  &  1  &   7168  & 2637.8  &  414.6  &  0.073  &  q \\
15  &  3403a &  52756.23  &  1  &   3541  & 2303.8  &  364.1  &  0.069  &  q \\
16  &  3403b &  52756.71  &  1  &   3158  & 1234.5  &  204.2  &  0.074  &  q \\
17  &  3404  &  52757.80  &  1  &   6977  & 1108.4  &  136.7  &  0.070  &  0 \\
18  &  3405  &  52758.99  &  1  &   6999  & 2021.7  &  194.9  &  0.055  &  0 \\
19  &  3407a &  52760.08  &  1  &   3439  & 1130.0  &  145.7  &  0.063  &  0 \\
20  &  3407b &  52760.56  &  2  &   3809  & 1152.9  &  145.1  &  0.071  &  0 \\
21  &  3408  &  52761.61  &  1  &   5214  & 1694.6  &  216.0  &  0.066  &  0 \\
22  &  3409  &  52762.75  &  1  &   6533  & 1733.9  &  256.1  &  0.067  &  0 \\
23  &  3410a &  52763.10  &  1  &   3418  & 1985.9  &  300.3  &  0.059  &  q \\
24  &  3410b &  52763.62  &  1  &   2769  & 2219.5  &  390.0  &  0.061  &  q \\
25  &  3411  &  52764.91  &  1  &   6755  & 2228.6  &  392.5  &  0.062  &  q \\
26  &  3412  &  52765.87  &  1  &   3398  & 3775.2  & 1083.3  &  0.047  &  q \\
27  &  3413  &  52766.57  &  1  &   2558  & 1569.8  &  392.9  &  0.143  &  q-bl \\
28  &  3414  &  52767.87  &  1  &   6784  & 1292.2  &  352.2  &  0.159  &  q-bl \\
29  &  3415  &  52768.55  &  1  &   2154  & 1307.5  &  336.5  &  0.149  &  q-bl \\
30  &  3416  &  52769.75  &  1  &   2433  & 1227.6  &  326.0  &  0.159  &  q-bl \\
31  &  3417  &  52770.53  &  2  &   3729  & 1533.1  &  387.0  &  0.140  &  q-bl \\
32  &  3418  &  52771.87  &  2  &   7672  & 1066.5  &  347.5  &  0.192  &  q-bl \\
33  &  3419  &  52772.74  &  1  &   6068  & 1000.1  &  368.6  &  0.214  &  q-bl \\
34  &  3420  &  52773.73  &  1  &   6757  &  994.0  &  365.3  &  0.212  &  q-bl \\
35  &  3421  &  52774.58  &  1  &   6702  & 1036.4  &  330.9  &  0.194  &  q-bl \\
36  &  3422  &  52775.63  &  1  &   6708  &  953.6  &  339.1  &  0.212  &  q-bl \\
37  &  3423  &  52776.68  &  1  &   6607  &  712.7  &  280.1  &  0.227  &  q-bl \\
38  &  3424  &  52777.67  &  1  &   6588  &  979.7  &  333.6  &  0.207  &  q-bl \\
39  &  3425  &  52778.53  &  1  &   6550  & 1012.4  &  319.8  &  0.192  &  q-bl \\
40  &  3426  &  52779.58  &  2  &   4873  & 1073.7  &  316.4  &  0.186  &  q-bl \\
41  &  3427  &  52780.60  &  1  &   3395  & 1079.9  &  317.4  &  0.184  &  q-bl \\
42  &  3428  &  52781.62  &  1  &   6805  & 1046.5  &  314.7  &  0.191  &  q-bl \\
43  &  3429  &  52782.70  &  1  &   3265  & 1093.2  &  299.9  &  0.173  &  q-bl \\
44  &  3430  &  52783.51  &  1  &   6529  & 1430.5  &  357.0  &  0.129  &  q-bl \\
45  &  3431  &  52784.57  &  1  &   6721  & 2082.5  &  409.6  &  0.078  &  q \\
46  &  3432  &  52785.48  &  1  &   6711  & 2158.8  &  432.0  &  0.083  &  q \\
47  &  3433  &  52786.35  &  1  &   6433  & 2517.7  &  597.7  &  0.067  &  q \\
48  &  3435  &  52788.51  &  2  &   7095  & 1606.6  &  240.0  &  0.056  &  q \\
49  &  3436  &  52789.26  &  1  &   3477  & 1908.4  &  327.2  &  0.057  &  q \\
50  &  3437  &  52790.24  &  2  &   3990  & 1781.1  &  296.6  &  0.055  &  q \\
51  &  3438  &  52791.63  &  1  &   5796  & 1717.7  &  272.1  &  0.061  &  q \\
52  &  3439  &  52792.39  &  2  &   6737  & 1755.4  &  296.6  &  0.056  &  q \\
53  &  3440  &  52793.61  &  1  &   3138  & 1785.1  &  286.8  &  0.052  &  q \\
54  &  3441  &  52794.56  &  1  &   7321  & 2558.6  &  636.5  &  0.030  &  q \\
55  &  3442  &  52795.38  &  2  &   3741  & 1587.5  &  236.5  &  0.050  &  q \\
56  &  3443  &  52796.27  &  1  &   7018  & 1785.1  &  307.2  &  0.060  &  q \\
57  &  3444  &  52797.58  &  1  &   7159  & 1806.0  &  325.6  &  0.066  &  q \\
58  &  3445  &  52798.54  &  1  &   3297  & 2647.5  &  675.5  &  0.029  &  q \\
59  &  3446  &  52799.46  &  1  &   2998  & 1561.3  &  233.7  &  0.048  &  q \\
60  &  3447  &  52800.73  &  2  &  11651  & 1528.1  &  215.4  &  0.035  &  q \\
61  &  3448  &  52801.91  &  1  &   4856  & 1619.8  &  260.7  &  0.055  &  q \\
62  &  3449  &  52802.96  &  1  &   4563  & 1614.5  &  272.1  &  0.057  &  q \\
63  &  3450  &  52803.57  &  1  &   6352  & 1539.1  &  259.6  &  0.058  &  q \\
64  &  3451  &  52804.62  &  1  &   6803  & 1607.7  &  298.8  &  0.069  &  q \\
65  &  3452  &  52805.47  &  1  &   5030  & 1577.6  &  278.3  &  0.063  &  q \\
66  &  3453  &  52806.63  &  1  &   3321  & 1361.0  &  187.3  &  0.032  &  0 \\
67  &  3454  &  52807.62  &  1  &   2431  & 1366.3  &  170.1  &  0.033  &  0 \\
68  &  3455  &  52808.64  &  1  &   5880  & 1311.8  &  162.4  &  0.032  &  0 \\
69  &  3456  &  52809.62  &  1  &   5203  & 1261.5  &  145.1  &  0.033  &  0 \\
70  &  3457  &  52810.54  &  1  &   5768  & 1137.5  &  101.8  &  0.033  &  0 \\
71  &  3458  &  52811.46  &  2  &   4235  & 1160.8  &  129.1  &  0.035  &  0 \\
72  &  3459  &  52812.67  &  1  &   4844  & 1132.5  &  120.9  &  0.038  &  0 \\
73  &  3461  &  52814.47  &  1  &   3930  & 1112.8  &  102.3  &  0.027  &  0 \\
74  &  3462  &  52815.84  &  1  &   3303  & 1003.0  &   77.0  &  0.030  &  0 \\
75  &  3463  &  52816.58  &  1  &   4889  &  978.9  &   75.7  &  0.030  &  0 \\
76  &  3464  &  52817.51  &  1  &   5704  & 1007.9  &   80.0  &  0.029  &  0 \\
77  &  3465  &  52818.55  &  1  &   4621  &  994.1  &   74.4  &  0.028  &  0 \\
78  &  3466  &  52819.34  &  2  &   2731  &  970.2  &   73.2  &  0.029  &  0 \\
79  &  3467  &  52820.43  &  1  &   2418  &  943.9  &   63.6  &  0.029  &  0 \\
80  &  3468  &  52821.42  &  1  &   2331  &  927.6  &   58.1  &  0.028  &  0 \\
81  &  3469  &  52822.53  &  1  &   4410  &  922.1  &   57.1  &  0.025  &  0 \\
82  &  3470  &  52823.48  &  1  &   4759  &  873.8  &   53.2  &  0.024  &  0 \\
83  &  3471  &  52824.45  &  1  &   3329  &  881.8  &   52.1  &  0.027  &  0 \\
84  &  3472  &  52825.37  &  1  &   2640  &  870.8  &   51.0  &  0.030  &  0 \\
85  &  3473  &  52826.42  &  1  &   3370  &  893.7  &   54.8  &  0.025  &  0 \\
86  &  3474  &  52827.41  &  1  &   3274  &  864.8  &   52.6  &  0.028  &  0 \\
87  &  3475  &  52828.37  &  1  &   6712  &  869.0  &   51.7  &  0.026  &  0 \\
88  &  3476  &  52829.28  &  2  &   5002  &  839.2  &   49.3  &  0.025  &  0 \\
89  &  3477  &  52830.19  &  2  &    949  &  863.9  &   52.8  &  0.027  &  0 \\
90  &  3478  &  52831.48  &  1  &   3371  &  860.5  &   51.2  &  0.025  &  0 \\
91  &  3481  &  52834.21  &  2  &   4408  &  838.0  &   49.3  &  0.023  &  0 \\
92  &  3484  &  52837.20  &  1  &   3492  &  834.4  &   47.4  &  0.024  &  0 \\
93  &  3487  &  52840.56  &  1  &   2137  &  837.5  &   49.9  &  0.027  &  0 \\
94  &  3490  &  52843.51  &  1  &   3179  &  816.3  &   52.9  &  0.026  &  0 \\
95  &  3493  &  52846.17  &  2  &   4941  &  792.2  &   45.0  &  0.024  &  0 \\
96  &  3499  &  52852.99  &  1  &   1770  &  749.2  &   41.0  &  0.027  &  0 \\
97  &  3502  &  52855.54  &  1  &   3696  &  722.0  &   37.5  &  0.022  &  0 \\
98  &  3509  &  52862.38  &  1  &   1766  &  668.4  &   35.1  &  0.029  &  0 \\
99  &  3512  &  52865.41  &  1  &   1952  &  630.1  &   31.9  &  0.031  &  0 \\
100 &  3516  &  52869.35  &  1  &   1463  &  600.6  &   31.1  &  0.032  &  0 \\
101 &  3522  &  52875.12  &  1  &   1655  &  569.6  &   28.3  &  0.028  &  0 \\
102 &  3524  &  52877.70  &  1  &    676  &  557.9  &   26.7  &  0.028  &  0 \\
103 &  3530  &  52883.29  &  1  &   3043  &  511.4  &   24.8  &  0.029  &  0 \\
104 &  3533  &  52886.57  &  1  &   1919  &  498.1  &   20.5  &  0.030  &  0 \\
105 &  3535  &  52888.93  &  1  &   2342  &  493.6  &   23.3  &  0.023  &  0 \\
106 &  3538  &  52891.43  &  1  &   2801  &  504.4  &   32.3  &  0.025  &  0 \\
107 &  3541  &  52894.05  &  1  &   1062  &  489.2  &   31.4  &  0.028  &  0 \\
108 &  3545  &  52898.72  &  1  &    953  &  426.2  &   26.3  &  0.032  &  0 \\
109 &  3547  &  52900.88  &  1  &    926  &  419.9  &   19.5  &  0.033  &  0 \\
110 &  3550  &  52903.99  &  1  &    769  &  392.3  &   19.6  &  0.033  &  0 \\
111 &  3552  &  52905.96  &  1  &    537  &  385.0  &   22.8  &  0.036  &  0 \\

\enddata
 
\tablenotetext{a}{RXTE day is the mission MET time in units of truncated days.}
\tablenotetext{b}{Midpoint of observation, $MJD=JD-2,400,000.5$.}  
\tablenotetext{c}{Source count rate in PCA at 2--35 keV, using PCU\#2.}  
\tablenotetext{d}{Source count rate in PCA at 7--35 keV, using PCU\#2.} 
\tablenotetext{e}{RMS source fluctuations at 2-35 keV, integrated over the range 0.1--10 Hz, and expressed as a fraction of the average source count rate.}
 
\label{tab:obs}

\end{deluxetable}

\clearpage

\newpage

\newpage

\end{document}